\documentclass[prb,aps,amsmath,showpacs,preprint]{revtex4}

\usepackage{graphicx}

\usepackage{dcolumn}

\usepackage{bm}

\begin{document}
\title{ Electronic structure of transition metal impurities in p-type ZnO} 
\author{L. Petit$^{1}$, T. C. Schulthess$^{1}$, A. Svane$^{2}$, Z. Szotek$^{3}$, W.M. Temmerman$^{3}$, and A. Janotti$^{1}$}
\affiliation{$^{1}$Computer Science and Mathematics Division, and Center for Computational Sciences, Oak Ridge National Laboratory, Oak Ridge, TN 37831, USA\\$^{2}$ Institute of Physics and Astronomy, University of Aarhus, 
DK-8000 Aarhus C, Denmark\\
$^{3}$ Daresbury Laboratory, Daresbury, Warrington WA4 4AD, UK}
\date{\today}

\begin{abstract}
The self-interaction corrected local spin-density approximation is used
to investigate the ground-state valency configuration of transition metal 
(TM=Mn, Co) impurities in p-type ZnO. Based on the total energy considerations, 
%
we find a stable localised TM$^{2+}$ configuration for a TM impurity in ZnO
if no additional hole donors are present. Our calculations indicate that the
(+/0) donor level is situated in the band gap, as a consequence of which the
TM$^{3+}$ becomes more favourable in p-type ZnO, where the Fermi level is
positioned at the top of the valence band. When co-doping with N, it
emerges that the 
carrier-mediated ferromagnetism can be realized in the scenario where
the N concentration exceeds the TM impurity concentration. If TM and N
concentrations are equal, the shallow acceptor levels introduced by N are
fully compensated by delocalised TM d-electrons.
\end{abstract}
\maketitle


Based on the possible interplay between electronic properties and spin-functionality, diluted magnetic semiconductors (DMS) are expected to play a major part in the development of the next generation of electronic devices.~\cite{wolf} The considerable challenge consists in designing  materials that remain ferromagnetic above room temperature. In Mn doped GaAs, where ferromagnetism is well established, the currently highest achieved Curie temperature~\cite{edmonds} is T$_C$=159 K. Ferromagnetism at even higher temperatures has been reported in some semiconductors (for a review, see for example Pearton {\it et al.}~\cite{pearton}), but doubts persist as to the carrier induced nature of the observed magnetic order. Recently, the prediction by Dietl {\it et al.}~\cite{dietl0} of room temperature ferromagnetism in p-type Zn$_{1-x}$Mn$_x$O has generated considerable research activity, both in theory and experiment. So far, no conclusive experimental evidence has emerged that could either confirm or disprove the prediction. Various experimental investigations of the magnetic order in Zn$_{1-x}$Mn$_x$O give contradictory results, ranging from spin glass behaviour~\cite{fukumura} and paramagnetism~\cite{tiwari} to ferromagnetism below~\cite{jung} or even above~\cite{sharma} room temperature. It has also been suggested that the observed ferromagnetism might be due to precipitates containing manganese oxides.~\cite{kim} The situation is quite similar for Zn$_{1-x}$Co$_x$O where there exists some experimental evidence in support of ferromagnetism,~\cite{ueda} whilst other experiments seem to suggest anti-ferromagnetic behaviour.\cite{mizokawa,risbud} The very latest experimental study, that we are aware of, finds no evidence for magnetic order (down to T=2 K), in either Zn$_{1-x}$Mn$_x$O or Zn$_{1-x}$Co$_x$O.~\cite{lawes}  

The theoretical description of magnetism in transition metal doped ZnO centers mainly on two different methodologies. First, there is the Zener model description by Dietl {\it et al.},~\cite{dietl0} where the divalent transition metal impurity (Mn$^{2+}$:d$^5$, Co$^{2+}$:d$^7$) provides a localized spin, and where a possible ferromagnetism originates from the RKKY-like interaction between the localized transition metal moments and delocalized hole carriers.
In Mn doped III-V semiconductors, such as Ga$_{1-x}$Mn$_x$As, this scenario is straightforwardly realized, as the substitution of Ga ([Ar]3$d^{10}$4$s^2$4$p^1$) by Mn ([Ar]3$d^5$4$s^2$) simultaneously introduces magnetic moments and hole carriers. 
In Mn doped ZnO on the other hand, the itinerant hole carriers need to be introduced through additional doping, for example by substituting some of the O atoms by N atoms, which then gives p-type Zn$_{1-x}$Mn$_x$O.
The second theoretical approach consists of describing the DMS in the framework of {\it ab initio} electronic structure calculations. Here the overlap of $d$-orbitals of the neighbouring transition metal impurities results in the formation of delocalized band states. The exchange interaction splits the majority and minority spin bands, and depending on the density of spin polarized states at the Fermi level, ferromagnetism based on the double exchange mechanism becomes theoretically possible.   
The magnetic properties of transition metal doped ZnO have been calculated from first principles, using a wide range of different implementations, such as the KKR-CPA,~\cite{sato} LMTO,~\cite{uspenskii} PAW,~\cite{sharma} and pseudopotential~\cite{spaldin,risbud} methods,
all based on the local spin density (LSD) approximation.

Both the Zener model description, and the LSD based {\it ab initio} calculations are in agreement regarding the crucial role played by additional hole carriers in the ferromagnetism of Zn$_{1-x}$(TM)$_x$O, while the electronic state of the TM impurity, and consequently the mechanism behind the long range magnetic interaction, differ qualitatively between the two descriptions. The Zener model description is based on the assumption that TM impurity states are localized atomic-like orbitals and the additional carriers always have the character of the host valence band. In the picture that emerges from calculations based on the LSD the TM impurity states form deep impurity bands, {\it i.e.}, they are itinerant but their character is always different from the host valence band. The description of the transition metal $d$-states, as either localized at the Mn/Co sites or band-like delocalized throughout the crystal, depends on the relative importance of the on-site correlations on one hand, and the kinetic energy on the other hand. The LSD approximation has been very successful in describing metallic bonding, but since it ignores the exchange and correlation effects beyond those of the homogeneous electron gas, it can not account for the on-site localization of correlated electrons by the local Coulomb repulsion.  A different way of looking at the inadequacy of the LSD approximation for describing $d$- and $f$-states is that it introduces an unphysical self-interaction of an electron with itself, which is insignificant for extended band states, but considerable for atomic-like states. In the self-interaction corrected (SIC)-LSD both localized and delocalized states are treated on an equal footing, by subtracting from the LSD energy functional a self-interaction contribution of each $d$-electron, thus enhancing its localized nature.~\cite{svane,perdew} Such a localization of a $d$-state at the impurity site gives rise to a gain in SIC (localization) energy, but simultaneously results in the loss of any possible band formation energy. The groundstate $d$-electron configuration of the transition metal impurity is determined from the global minimum of the total energy.~\cite{svanegunnar} 
By studying various valency configurations of the transition metal ion, realized when treating some of the $d$-electrons as localized and allowing the remainder to hybridize, one can find both the global energy minimum and the configuration of localized orbitals. Here we assume that the groundstate spin and orbital moments of the TM $d$-orbitals are governed by the Hund's rules. 

ZnO crystallizes in the hexagonal wurtzite structure (with lattice constants a$_0$=3.25~{\AA}  
and c$_0$=5.21~\AA). Its wide band gap, in the near UV range, makes it a candidate for optoelectronic applications that
rely on short wavelength light emitting diodes. Another advantage in dealing with ZnO is that it is abundant, low cost and
environmentally friendly. For pure ZnO, we find that applying SIC-LSD, namely treating the Zn 3$d$ electrons as localized, results in an energy gap of E$_g$=3.7 eV, as compared to the smaller LSD value, E$_g$=1.8 eV, and in relatively good agreement with the experimental value, E$_g$=3.4 eV. Using the pulsed-laser deposition technique, Fukumura {\it  et al.}~\cite{fukumura1} were able to fabricate epitaxial thin films of DMS Zn$_{1-x}$Mn$_x$O (x $\leq$ 0.35), indicating a high solubility of Mn in the ZnO matrix and showing that the Mn ions occupy the Zn sites without changing the wurtzite structure. 
It was similarly shown that a large amount of Co can be substituted for Zn in ZnO, without any impurity phase appearing in wurtzite structure.~\cite{jayaram}

\begin{figure}[h]
\includegraphics[scale=0.30,angle=-90,clip]{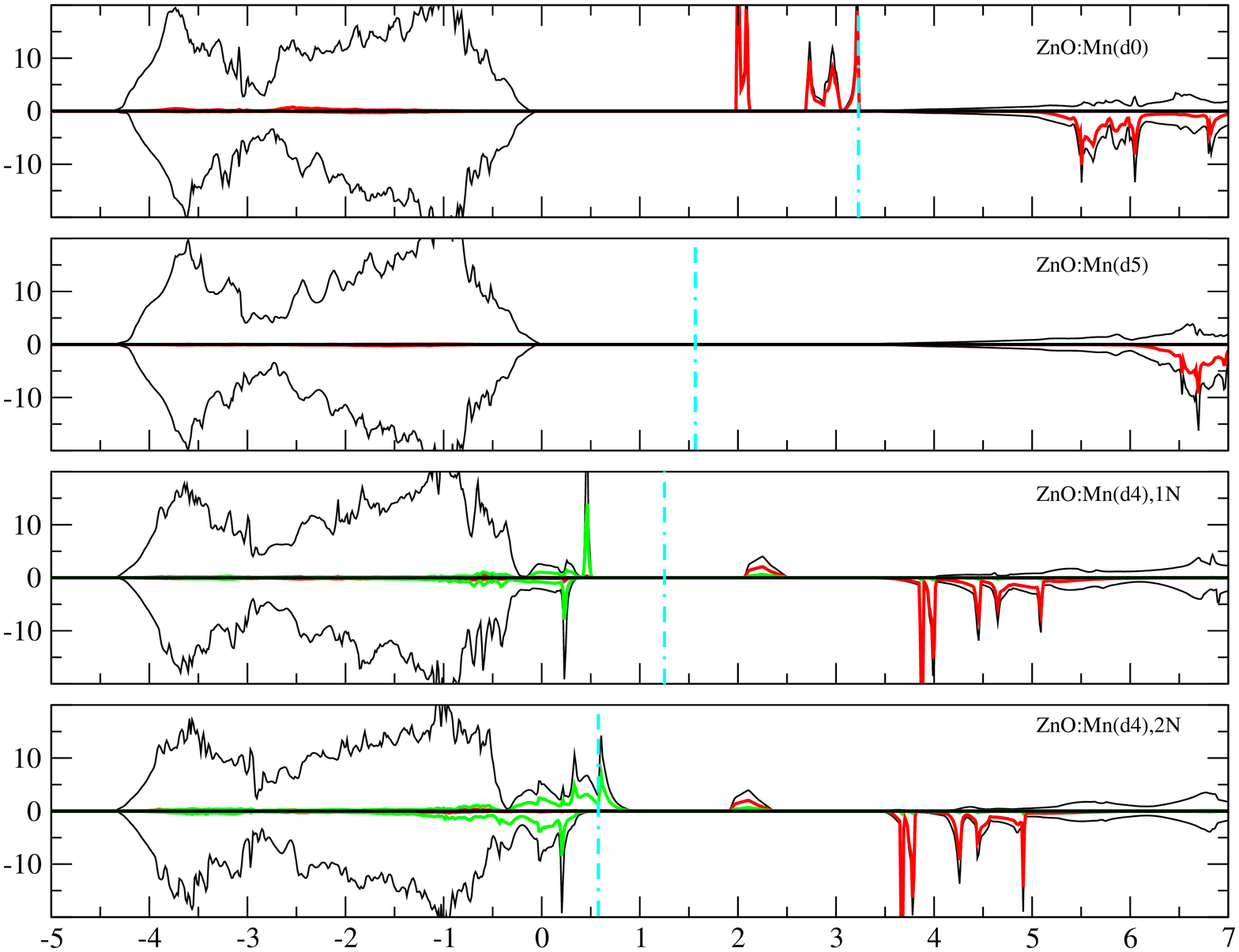}
\caption{Total DOS as a function of energy, in states per eV, of Zn$_{15/16}$TM$_{1/16}$O$_{1-y}$N$_y$: a) LSD configuration, y=0, b) TM$^{2+}$ configuration, y=0, c) TM$^{3+}$ configuration, y=1/16, d) TM$^{3+}$ configuration, y=2/16. 
The black, red, and green lines represent the total, TM $d$-projected, and N $p$-projected densities of states respectively. The energy is given relative to the VBM, with the Fermi level indicated by the light blue dash-dotted line.} 
\label{znmno}
\end{figure}

In our calculations, Zn$_{1-x}$TM$_x$O is realized by substituting a single Zn by either Mn or Co, in a (2x2x2) supercell consisting of 16 ZnO formula units.
In Fig. \ref{znmno}a, we show the density of states (DOS) of Mn doped ZnO, as calculated within the LSD approximation, i.e., when all the $d$-states are treated as band states. The O $p$-states make up most of the broad band situated below 0 eV. The exchange splitting separates the $d$-manifold (thick red line) into well defined majority and minority bands, with the Fermi level situated at the top of the completely filled majority band. The DOS of the LSD scenario for Co doping (not shown) is quite similar to Fig. \ref{znmno}a, except for the fact that the two additional $d$-electrons are now accommodated in the minority band. Based on the position of the Fermi level, it has been suggested that Zn$_{1-x}$Co$_x$O, but not Zn$_{1-x}$Mn$_x$O, is ferromagnetic, with the magnetic order being mediated by the double exchange mechanism.~\cite{sato} In contrast, calculations by Spaldin~\cite{spaldin} seem to indicate that ferromagnetism is unlikely to occur in either compound, unless carriers are added. 
When treating the transition metal $d$-states as localized, within the SIC-LSD approximation, 
an altogether different picture emerges, as can be seen from Fig. \ref{znmno}b. The SIC localizes 
the TM $d$-electrons in atomic-like orbitals and consequently shifts the corresponding $d$-states below the 
valence band, giving rise to a wide energy band gap. 

In Table I, the total energies for Zn$_{15/16}$Mn$_{1/16}$O (row 2), and Zn$_{15/16}$Co$_{1/16}$O (row 6) have been calculated for three different localized scenarios (columns 2, 3, and 4), as well as the LSD configuration (column 5). We find that the global energy minima are obtained when treating all the TM impurity $d$-states in ZnO as localized, where the preferred configurations are respectively Mn$^{2+}$, with five localized $d$-electrons, and Co$^{2+}$, with seven localized $d$-electrons. For both dopants the energy difference between the divalent and LSD scenarios is more than 3 eV, indicating that $d$-electron localization results in an overall gain in SIC energy that far outweighs any corresponding loss in hybridization energy. The divalent groundstate configuration is in agreement with experimental observation both for Mn,~\cite{dorain} and Co.~\cite{wi}
Given the fact that in the divalent scenario of Fig. \ref{znmno}b there are no available hole carriers to mediate the interaction between the localized spins, we conclude that neither Zn$_{1-x}$Mn$_{x}$O nor Zn$_{1-x}$Co$_{x}$O will be ferromagnetic.   

The influence of p-doping Zn$_{15/16}$TM$_{1/16}$O, by substituting one O with one N atom, is evident from Table I. As can be seen from rows 3 and 7, this has a major effect on the groundstate valency configuration of the TM impurity, as the global energy minimum is now obtained for the trivalent TM$^{3+}$ state. Nitrogen, having one $p$-electron less than O, acts as an acceptor when introduced into ZnO, with the corresponding acceptor level situated near the top of the valence band,~\cite{yamamoto} (Fig. \ref{znmno}c). Since one of the previously localized $d$-states becomes delocalized in the TM$^{3+}$ configuration, the corresponding electron charge transfers into the N acceptor state, leaving behind an empty $d$-band, situated at approximately 2 eV above the valence band maximum (VBM) for the Mn doped case shown in Fig. \ref{znmno}c. Again the situation is very similar for Zn$_{15/16}$Co$_{1/16}$O$_{15/16}$N$_{1/16}$, where the empty $d$-state is situated at approximately 1.5 eV above the VBM. The energy gain, resulting from the charge transfer and hybridization in the TM$^{3+}$ configuration, is obviously large enough to overcome the corresponding loss in SIC (localization) energy with respect to the TM$^{2+}$ configuration. The Fermi level is situated in the gap above the completely filled acceptor states, indicating that the compensation is complete, and that there are no carriers left to mediate the magnetic order.

In order to make the connection between the localization-delocalization picture, and the terminology used in describing doping in semiconductors, we need to analyze the ionization levels
of the TM impurities. To avoid confusion we introduce a notation that is better suited for describing neutral and charged impurities, i.e. (ZnO:TM$^v$,q), where {\it v} and q refer respectively to the valency and the charge state of the impurity. In the neutral charge state the substitutional TM impurity can either assume the divalent configuration, TM$^{2+}$ $\equiv$ (ZnO:TM$^{2+}$,0), which does not introduce states in the gap, or the trivalent configuration, TM$^{3+}$ $\equiv$ (ZnO:TM$^{3+}$,0), which is about 0.5-1.5 eV higher in energy (see Table I, columns 2 and 3), and has an occupied electron $d$ state in the gap. This electron can be transferred to the conduction band, depending on the Fermi level position, ionizing the TM impurity into the charge state (ZnO:TM$^{3+}$,+).
Thus, with respect to substitutional Co or Mn in the Zn$^{2+}$O$^{2-}$ matrix, the delocalization process, (ZnO:TM$^{2+}$,0) $\rightarrow$  (ZnO:TM$^{3+}$,+) + e$^-$, results in a deep donor level (+/0) in the band gap, which can be illustrated by calculating the formation energy $E_{f}$ of the neutral and the positively charged TM impurity using~\cite{walle}
\begin{equation}
E_f({\rm ZnO:TM}^{v},{\rm q})=E_{tot}({\rm ZnO:TM}^{v},{\rm q})-E_{tot}({\rm ZnO})-\mu_{\rm Zn}+\mu_{\rm TM}+{\rm q} \epsilon_F .
\label{Eform}
\end{equation}
Here $\mu_{\rm Zn}$ and $\mu_{\rm TM}$ are the respective chemical potentials, and $\epsilon_F$ is the Fermi level (with respect to VBM).
Although the absolute formation energies depend on the chemical potentials, here we will discuss the relative values that depend only on the total energy differences between the various TM valency configurations. The total energy of the positive charge state $E_{tot}({\rm ZnO:TM^{3+}},+)$ is estimated by subtracting the 
one-electron energy of the gap state from the total energy of the neutral (ZnO:TM$^{3+}$,0) configuration.~\cite{E+} The relative formation energies as a function of the Fermi energy are schematically plotted in Fig. 2. For the charged impurity configuration
(ZnO:TM$^{3+}$,+) the formation energy is determined by the position of the Fermi level, as follows from Eq. (1), and as indicated by the skew line.  The neutral charge configurations (ZnO:TM$^{3+}$,0) and (ZnO:TM$^{2+}$,0) are indicated by the horizontal dashed and solid lines respectively. The donor transition level (+/0) is defined by the Fermi energy above which the TM impurity is in the divalent configuration  and below which it is in the positively charged trivalent configuration. Since the neutral charged (ZnO:TM$^{3+}$,0)
has a gap state in the high energy part of the band gap, and since it is situated only 0.5-1.5 eV above the neutral (ZnO:TM$^{2+}$,0) configuration, the donor transition level (+/0), is also situated in the gap.  It is interesting to note that in the SIC-LSD picture, the doping of TM impurities into ZnO results in the donor level (+/0) being situated in the gap, as is also found
experimentally,\cite{madelung} as a result of two competing valence configurations, (ZnO:TM$^{2+}$,0) and (ZnO:TM$^{3+}$,+). Without N codoping, in both Zn$_{15/16}$Mn$_{1/16}$O and Zn$_{15/16}$Co$_{1/16}$O, the Fermi level is situated above the donor level (respectively columns 7 and 6 in Table I), which explains the stability of the localized TM$^{2+}$ configuration. 
With N codoping (rows 3, 4, and 7, 8), the additional acceptor state lowers $\epsilon_F$, energetically favoring the TM$^{3+}$ configuration.
This is in contrast to Dietl's Zener model description, where it is assumed that the groundstate configuration remains Mn$^{2+}$, based on the argumentation that the (+/0) level is situated below the VBM.~\cite{dietl2}

From the SIC-LSD calculations it follows that, the band description of the $d$-states does not fully account for the correct electronic structure of the Zn$_{1-x}$TM$_x$O ground state, even when codoped with N. But also the Zener model, with hole carriers mediating the magnetic interaction between localized spins residing on TM$^{2+}$ ions, is not a true representation of the groundstate, due to the fact that the TM(+/0) donor level is situated above the N acceptor levels which therefore are fully compensated by the delocalized $d$-electron. There are no hole carriers in either Zn$_{15/16}$Mn$_{1/16}$O$_{15/16}$N$_{1/16}$ or Zn$_{15/16}$Co$_{1/16}$O$_{15/16}$N$_{1/16}$ and this is why they cannot be ferromagnetic.   
Though we only investigated the electronic structure of ZnO, when codoped with both a single TM impurity and a single N, it seems plausible that increasing the concentrations of the dopants, [TM] and [N], will not change the overall picture of fully compensated acceptor states, as long as [TM]=[N]. However, the qualitative picture changes considerably if we increase the relative amount of N impurities, i.e., if [N]$>$[TM]. Substituting two of the O atoms by N, but with a single TM impurity in the 32 atom ZnO supercell, we find that, contrary to what one might expect, overdoping with N does not result in a further delocalization, i.e., a transition from TM$^{3+}$ $\longrightarrow$ TM$^{4+}$ + e$^-$. From Table I we see that the TM$^{3+}$ configuration gives the lowest total energy, both for Zn$_{15/16}$Mn$_{1/16}$O$_{14/16}$N$_{2/16}$ (column 3, row 4), and Zn$_{15/16}$Co$_{1/16}$O$_{14/16}$N$_{2/16}$ (column 3, row 8). The additional N impurity is thus not compensated by TM $d$-electrons and, from the corresponding DOS (Fig. \ref{znmno}d), we find that this results in an impurity band at the top of the valence band, which is now only partially filled. It is most noticeable that, since these hole states coexist simultaneously with the localized spins on the TM$^{3+}$ impurities, hole mediated ferromagnetism is now theoretically possible.
On the other hand, since codoping with N has revealed itself to be a rather difficult undertaking,~\cite{joseph} the condition [N]$>$[TM] constitutes a considerable hurdle with regard to actually synthesizing these DMS.


In summary, we have studied the electronic structure and different valency configurations 
of Co and Mn impurities in p-type ZnO using the SIC-LSD {\it ab initio} method. From total energy considerations we find that the TM $d$-states remain localized if no additional hole donors are present. The TM$^{3+}$ becomes more favourable in $p$-type ZnO, 
which leads us to concolude that carrier-mediated ferromagnetism can theoretically be realized with 
the N co-doping, when the latter exceeds the concentration of the TM impurities.

This work was supported in part by the Defense Advanced Research Project Agency and by the Division of Materials Science and Engineering, US Department of Energy. Oak Ridge National Laboratory is managed by UT-Batelle, LLC, for the US Department of Energy under Contract No. DE-AC05-00OR22725. The calculations were carried out at the Center for Computational Sciences at Oak Ridge National Laboratory, and at the Danish Center for Scientific Computing. 


\newpage

\begin{figure}[h]
\includegraphics[scale=0.40,angle=0,clip]{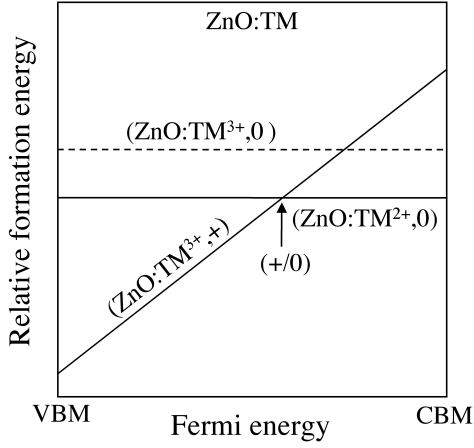}
\caption{Schematic plot of the formation energy as a function of Fermi energy for ZnO:TMn illustrating the link between the localization-delocalization picture and
the transition metal (TM) donor ionization level in the band gap. Only the (+/0) between the positively charged TM$^{3+}$ and the neutral TM$^{2+}$ valence configurations
is indicated. }
\label{principle}
\end{figure}

\newpage

\begin{table}[b]
\caption{Columns 2-5, total energy (in eV) of Zn$_{15/16}$TM$_{1/16}$O$_{1-y}$N$_y$ for TM$^{2+}$, TM$^{3+}$, TM$^{4+}$, and LSD configurations, with TM=Mn or Co, and y=0, 1/16, and 2/16. Column 6, the Fermi level (with respect to the VBM) as obtained in the TM$^{3+}$ configuration. Column 7, the donor level (+/0) with respect to the VBM (given in eV).  }
\label{localize}
\begin{ruledtabular}
\begin{tabular}{|l|l|l|l|l|l|l|} & Mn$^{2+}(d^5) $ & Mn$^{3+}(d^4)$ & Mn$^{4+}(d^3)$  & Mn$^{7+}(d^0)$& $\epsilon_F$ (d$^4$)&Mn(+/0)\\
\hline
Zn$_{15/16}$Mn$_{1/16}$O &{\bf -113.132}&-112.606&-112.031&-110.115&3.22&2.70 \\
Zn$_{15/16}$Mn$_{1/16}$O$_{15/16}$N$_{1/16}$ &-111.398&{\bf -113.166}&-112.581&&1.28&3.04 \\
Zn$_{15/16}$Mn$_{1/16}$O$_{14/16}$N$_{2/16}$ &-110.201&{\bf -111.398}&-111.349&&0.60&1.80 \\
\hline
\hline & Co$^{2+}(d^7) $ & Co$^{3+}(d^6)$ & Co$^{4+}(d^5)$  & Co$^{9+}(d^0)$& $\epsilon_F$ (d$^6$)&Co(+/0) \\
\hline
Zn$_{15/16}$Co$_{1/16}$O  &{\bf -114.036}&-113.397&-112.703&-108.378&3.51&2.87 \\
Zn$_{15/16}$Co$_{1/16}$O$_{15/16}$N$_{1/16}$ &-112.308&{\bf -113.900}&-113.247&&1.06&2.64\\
Zn$_{15/16}$Co$_{1/16}$O$_{14/16}$N$_{2/16}$ &-110.282&{\bf -112.010}&-111.788&&0.54&2.27\\
\end{tabular}
\end{ruledtabular}
\end{table}

\begin{references}
\bibitem{E+}This approximation is valid when the effects of the atomic relaxations are small and when the one-particle state is not doubly occupied.
\end{references}
\end{document}